\documentclass[aps, prd, twocolumn, amsmath, floats,floatfix, superscriptaddress, nofootinbib,
showpacs]{revtex4}

\usepackage{amssymb}
\usepackage{amsmath}
\usepackage{verbatim}
\usepackage{mathrsfs}
\usepackage{amsfonts}
\usepackage{latexsym}
\usepackage{epsfig}
\usepackage{color}
\usepackage{graphicx,subfigure}
\usepackage{units}
\usepackage{slashbox}
\usepackage{envmath}
\usepackage{natbib}

\begin{document}

\title{Quasi-stationary solutions of self-gravitating scalar fields
  around black holes} 

\author{Nicolas Sanchis-Gual}
\affiliation{Departamento de
  Astronom\'{\i}a y Astrof\'{\i}sica, Universitat de Val\`encia,
  Dr. Moliner 50, 46100, Burjassot (Val\`encia), Spain}

\author{Juan Carlos Degollado} 
\affiliation{Departamento de Ciencias Computacionales,
Centro Universitario de Ciencias Exactas e Ingenier\'ia, Universidad de Guadalajara\\
Av. Revoluci\'on 1500, Colonia Ol\'impica C.P. 44430, Guadalajara, Jalisco, Mexico}

\author{Pedro J. Montero} 
\affiliation{Max-Planck-Institute f{\"u}r Astrophysik, Karl-Schwarzschild-Str. 1, 85748, Garching bei M{\"u}nchen, Germany}

\author{Jos\'e A. Font}
\affiliation{Departamento de
  Astronom\'{\i}a y Astrof\'{\i}sica, Universitat de Val\`encia,
  Dr. Moliner 50, 46100, Burjassot (Val\`encia), Spain}
\affiliation{Observatori Astron\`omic, Universitat de Val\`encia, C/ Catedr\'atico 
  Jos\'e Beltr\'an 2, 46980, Paterna (Val\`encia), Spain}


\date{December 2014}


\begin{abstract}  

Recent perturbative studies have shown the existence of long-lived, quasi-stationary configurations 
of scalar fields around black holes. In particular, such configurations have been found to survive for 
cosmological timescales, which is a requirement for viable dark matter halo models in galaxies based 
on such type of structures. In this paper we perform a series of numerical relativity simulations of 
dynamical non-rotating black holes surrounded by self-gravitating scalar fields. We solve numerically 
the coupled system of equations formed by the Einstein and the Klein-Gordon equations under the 
assumption of spherical symmetry using spherical coordinates. Our results  confirm the existence of 
oscillating, long-lived, self-gravitating scalar field configurations around non-rotating black holes in 
highly dynamical spacetimes with a rich scalar field environment. Our numerical simulations are 
long-term stable and allow for the extraction of the resonant frequencies to make a direct comparison 
with results obtained in the linearized regime. A byproduct of our simulations is the existence of a 
degeneracy in plausible long-lived solutions of Einstein equations that would induce the same motion 
of test particles, either with or without the existence of quasi-bound states.
 
\end{abstract}




\maketitle

\section{Introduction}\label{sec:introduction}

Scalar fields play an important role in different areas of theoretical physics. For instance, recent models 
for the evolution of the universe, dark energy and string theory require the existence of ultra-light scalar 
degrees of freedom \cite{Arvanitaki:2009fg, Arvanitaki:2010sy, Kamionkowski:2014zda}. In a cosmological
context, scalar fields have been suggested as the constituents of dark
matter halos in galaxies \cite{Hwang:1996xd, Matos:1999et, Matos:2000ss, Matos:2000ng, Marsh:2010wq, Hu:2000ke}. There
is large observational evidence that points to the presence of supermassive black
holes in the centers of most nearby galaxies. Furthermore, the evolution of these black holes is
directly connected to their host galaxy evolution \cite{Kormendy01}. 
It has been suggested that in order to consider scalar fields as plausible candidates for dark matter halos 
they must survive for cosmological time scales in the presence of black holes \cite{Burt:2011pv}. 

There are several studies in the literature investigating the
dynamics of scalar fields around black holes within the linearized
regime, i.e.~in the test field regime in which the black hole spacetime is taken as a
background where the scalar field evolves \cite{Witek:2012tr,Barranco:2012qs, Barranco:2013rua}. 
One of the key results of linearized studies concerning massive scalar fields surrounding
stationary black holes is that as a result of the presence of a potential well due to the mass term,
two types of modes are possible depending on the boundary conditions at infinity: no-normalizable
quasi-normal modes and quasi-bound states, which decay at infinity and thus are localized around
the black hole. If the black hole is spinning  the latter class can yield exponentially growing
modes \cite{Detweiler:1980uk, Dolan:2007mj} and to hairy black hole solutions \cite{Herdeiro:2014goa}.

Quasi-bound states are scalar field configurations around black holes that may be very long lived 
\cite{Barranco:2013rua}. Many of their characteristics can be derived in the test field
approximation. For instance, it is possible to determine their complex frequency by imposing 
ingoing boundary conditions at the black hole horizon and an asymptotically decay behaviour at 
infinity. The problem of obtaining the frequencies of the quasi-bound states reduces to an eigenvalue problem which can be solved in several efficient ways (see e.g.~\cite{Berti:2009kk,Konoplya:2011qq}). The real part represents the oscillation frequency and the imaginary part gives their rate of decay. 

While the test field approximation is valid as long as the energy
content of the field is small it will break down eventually as the
energy increases. At some point, the backreaction of the scalar field onto the spacetime
dynamics becomes non-negligible and non-linear simulations of
self-gravitating scalar field configurations are necessary. There are recent studies in this 
direction, most notably~\cite{Okawa:2014nda}, where a fully 3D evolution of scalar fields around black holes was performed. In such work the authors made a thorough study of scalar field configurations 
around stationary and rotating black holes finding quasi-bound states when the amount of scalar 
field contributes to some fraction of the mass of the black hole. Here we focus on the cases in 
which the mass of the scalar field  cloud may be {\it greater} than the mass of the black hole and 
therefore constitutes a strong deviation from the linearized regime. Stated in a different way, our 
setup involves a large scalar field environment in a highly dynamical spacetime. These scenarios 
may be framed as the final stage of a violent process such as the bosenova 
explosion \cite{Yoshino:2012kn} or as a Kerr hairy solution in which the black hole has lost its 
spin. 

The use of spherical symmetry allows us to follow accurately the evolution of the system
and describe the development of a single mode without interference effects from other modes. 
Furthermore, the significantly long runs we can afford ($t\sim10^{5}M$, where $M$ is the bare mass of the black hole) allow us to obtain very precise figures
for the frequencies of the bound states.  A comment on orders of magnitude becomes necessary 
at this point as it should be noted that the frequency of the quasi-bound states for very small 
values of the scalar field mass $M\mu$
compatible with dark matter models, is of the order of $M\omega \sim 10^{-6}$ in geometrized units 
\footnote{In physical units this corresponds to $\hbar\mu\sim 10^{-24}$ eV \cite{Matos:2000ss,Lundgren:2010sp}.}. 
According to our results, in order to capture accurately the values of the corresponding frequencies one would have to evolve the system 
at least for $t\sim10^{10}M$. Such timescale, however, cannot be reached even at the linear level. 
For this reason we will consider values of $\mu$ in the range from $0.08$ to $0.3$ as we discuss below.

For this work we have coupled the Klein-Gordon
equation to the Einstein equations taking advantage of the computational framework recently 
proposed by Montero and Cordero-Carri\'on \cite{Montero:2012yr} in which a second-order 
partially-implicit Runge-Kutta (PIRK) method is applied to the Baumgarte-Shapiro-Shibata-Nakamura (BSSN) formulation of the Einstein equations. This approach has been shown to lead to long-term 
stable numerical simulations of both vacuum and non-vacuum (with fluids) spacetimes using spherical coordinates without the need for a regularization algorithm at the origin. 

The paper is organized as follows:
Section~\ref{sec:formalism} describes the formulation of the coupled
Einstein-Klein-Gordon system and the construction of the initial
data. Section~\ref{sec:NumImpl} focuses on the time integration of
the system of evolution equations. Our results are presented and discussed in
Section~\ref{sec:num_results}. Finally, we end with a summary of our findings in
Section~\ref{sec:conclusions}. 
In the following sections Greek indices $\alpha, \beta, \ldots$ run over spacetime indices, while 
Latin indices $i, j, \ldots$ run over space indices only. Throughout this article we use geometrized units $c=G=1$.

\section{Basic equations}\label{sec:formalism}

We investigate the dynamics of a self-gravitating scalar field
configuration around a black hole by solving numerically the coupled
Einstein-Klein-Gordon system 
\begin{equation}
 R_{\alpha\beta}-\frac{1}{2}g_{\alpha\beta}R=8\pi T_{\alpha\beta} \ ,
\label{eq:Einstein}
\end{equation}
with matter content given by the stress energy tensor
\begin{equation}
 T_{\alpha\beta} = \partial_{\alpha}\Phi \partial_{\beta}\Phi-\frac{1}{2}g_{\alpha\beta}\left(\partial^{\sigma}\Phi\partial_{\sigma}\Phi+
\mu^2\Phi^2 \right)\ .
\label{eq:tmunu}
\end{equation}
The conservation of this stress-energy tensor implies that the field obeys
the Klein-Gordon equation
\begin{equation}
 \Box \Phi-\mu^2\Phi=0 \ ,
\label{eq:KG}
\end{equation}
where the D'Alambertian operator is defined by $\Box:=
(1/\sqrt{-g})\partial_{\alpha}(\sqrt{-g}g^{\alpha\beta}\partial_{\beta})$. We
follow the convention that $\Phi$ is dimensionless and $\mu$ has
dimensions of (length)$^{-1}$.

\subsection{Einstein's equations in spherical symmetry}

We assume that the spacetime $\cal{M}$ can be foliated by a family of spatial slices $\Sigma_t$ 
that coincide with level surfaces of a coordinate time $t$.   
We denote the future-pointing unit normal on $\Sigma_t$ with $n^{\alpha}$ and write the spacetime metric $g_{\alpha\beta}$ as
\begin{eqnarray} \label{metric}
ds^2 & = & g_{\alpha\beta} dx^\alpha dx^\beta \nonumber \\ 
& = & - \alpha^2 dt^2 + \gamma_{ij} (dx^i + \beta^i dt)(dx^j + \beta^j dt),
\end{eqnarray}
where $\alpha$ is the lapse function, $\beta^i$ the shift vector, and $\gamma_{ij}$ the spatial metric induced on $\Sigma$,
\begin{equation} \label{spatial_metric_def}
\gamma_{\alpha \beta} = g_{\alpha \beta} + n_{\alpha} n_{\beta}.
\end{equation}
 In terms of the lapse and the shift, the normal vector $n^{\alpha}$ can be expressed as
\begin{equation} \label{normal}
n^{\alpha} = (1/\alpha, - \beta^i/\alpha)~~~\mbox{or}~~~n_{\alpha} = (-\alpha,0,0,0) .
\end{equation}

We adopt a conformal decomposition of the spatial metric $\gamma_{ij}$ 
\begin{equation} \label{conformal_decomposition}
\gamma_{ij} = e^{4 \chi} \hat \gamma_{ij},
\end{equation}
where $\psi = e^\chi = (\gamma/\hat \gamma)^{1/12}$ is the conformal
factor, $\hat \gamma_{ij}$ the conformally related metric and $\hat
\gamma$ its determinant.

Under the assumption of spherical symmetry the line element may be
written as
\begin{equation}
 ds^2 = e^{4\chi } (a(t,r)dt^2+ r^2\,b(t,r)  d\Omega^2) \ ,
\end{equation}
with $d\Omega^2 = \sin^2\theta d\varphi^2+d\theta^2$ being the solid angle element
and $a(t,r)$ and $b(t,r)$ the metric functions.

We set the value of $\hat\gamma$ at $t=0$ as that of the determinant of the flat metric in spherical coordinates $\mathring\gamma$. The evolution equations for the conformal factor and for the conformal metric components take the form
\begin{eqnarray}
\label{eq:X}\partial_{t}X&=&\beta^{r}\partial_{r}X-\frac{1}{3}X\hat{\nabla}_{m}\beta^{m}+\frac{1}{3}X\alpha K \ ,\\
\label{eq:a}\partial_{t}a&=&\beta^{r}\partial_{r}a+2a\partial_{r}\beta^{r}-\frac{2}{3} a\hat{\nabla}_{m}\beta^{m}-2\alpha aA_{a} \ ,\,\, \\
\label{eq:b}\partial_{t}b&=&\beta^{r}\partial_{r}b+2b\frac{\beta^{r}}{r}-\frac{2}{3} b\hat{\nabla}_{m}\beta^{m}-2\alpha bA_{b} \ ,
\end{eqnarray}
where $X \equiv e^{-2\chi}$, $K$ is the trace of the extrinsic curvature,  
$\hat{\nabla}_{m}\beta^{m}$ is the divergence of the shift vector $\beta^{i}$, $\hat{A}_{ij}$ is the traceless 
part of the conformal extrinsic curvature, and 
\begin{equation}
\label{eq:A_a}
	A_{a} \equiv \hat{A}^{r}_{r}\;,\;\;\;\;\;\;
A_{b} \equiv \hat{A}^{\theta}_{\theta} \ . 
\end{equation}
The evolution equation for $K$ is
\begin{align}
\label{eq:K}
	\partial_{t} K  & =  \beta^{r} \partial_{r}K - \nabla^{2}\alpha +
\alpha(A_{a}^{2} + 2A_{b}^{2} + \frac{1}{3}K^{2}) \nonumber \\
    & + 4\pi\alpha(\rho+S_{a}+2S_{b}) \ .
\end{align}

Next, the evolution equation for the independent component of the traceless 
part of the conformal extrinsic curvature, $A_a$, is given by
\begin{align}
\label{eq:Aa}
	\partial_{t} A_{a} & = \beta^{r}\partial_{r}A_{a} -
\left(\nabla^{r}\nabla_{r}\alpha - \frac{1}{3}\nabla^{2}\alpha\right)
+ \alpha\left(R^{r}_{r} - \frac{1}{3}R\right) \nonumber \\
	& + \alpha K A_{a} - \frac{16}{3}\pi \alpha(S_a - S_b) \ ,
\end{align}

where $R^{r}_{r}$ is the mixed radial component of the Ricci tensor and $R$ is its 
trace.

Finally, the evolution equation for $\hat {\Delta}^{r}$, the radial component 
of the additional BSSN variables \cite{Alcubierre:2010is} 
$\hat {\Delta}^{i} = \hat{\gamma}^{mn} \hat {\Delta}^{i}_{mn}$ with 
$\hat {\Delta}^{a}_{bc} = \hat{\Gamma}^{a}_{bc}-\mathring{\Gamma}^{a}_{bc}$, is
given by 
\begin{align}
\label{eq:Deltar}
	\partial_{t}\hat {\Delta}^{r} & = \beta^{r}\partial_{r}\hat{\Delta}^{r} 
- \hat{\Delta}^{r}\partial_{r}\beta^{r} + \frac{1}{a}\partial^{2}_{r}\beta^{r} 
+ \frac{2}{b}\partial_{r}\left(\frac{\beta^r}{r}\right) \nonumber \\
 &+ \frac{\sigma}{3}\left(\frac{1}{a}\partial_{r}(\hat{\nabla}_m\beta^{m}) 
+ 2\hat{\Delta}^{r}\hat{\nabla}_m\beta^{m}\right) \nonumber \\
 &- \frac{2}{a}(A_{a}\partial_{r}\alpha + \alpha\partial_{r}A_{a}) \nonumber \\
 &+ 2\alpha\left(A_{a}\hat{\Delta}^{r} - \frac{2}{rb}(A_{a}-A_{b})\right)
\nonumber \\
 &+ \frac{2 \alpha}{a} \left[\partial_{r}A_{a} - \frac{2}{3}\partial_{r}K 
+ 6A_{a}\partial_{r}\chi \right. \nonumber \\
 & \left. +(A_{a}-A_{b})\left(\frac{2}{r}+\frac{\partial_{r}b}{b}\right) 
- 8\pi j_{r} \right].
\end{align}

The matter source terms $\rho$, $S_{a}$, $S_{b}$ and $j_r$ appearing in the previous equations are components of the energy-momentum
tensor \eqref{eq:tmunu} given by
\begin{eqnarray}
\rho&:=&n^{\alpha}n^{\beta}T_{\alpha\beta}=\frac{1}{2}\biggl(\Pi^{2}+\frac{\Psi^{2}}{ae^{4\chi}}\biggl)+\frac{1}{2}\mu^{2}\Phi^{2} \label{eq:rho}\ ,\\
j^{r}&:=&-P^{r\alpha}n^{\beta}T_{\alpha\beta}=-\Pi\Psi,\\
S_{a}&:=&T^{r}_{r}=\frac{1}{2}\biggl(\Pi^{2}+\frac{\Psi^{2}}{ae^{4\chi}}\biggl)-\frac{1}{2}\mu^{2}\Phi^{2} \ ,\\
S_{b}&:=&T^{\theta}_{\theta}=\frac{1}{2}\biggl(\Pi^{2}-\frac{\Psi^{2}}{ae^{4\chi}}\biggl)-\frac{1}{2}\mu^{2}\Phi^{2} \ ,
\end{eqnarray}
where $\Pi$ and $\Psi$ are defined below.
The Hamiltonian and momentum constrains are given by the following two equations that we compute 
to monitor the accuracy of the numerical evolutions: 
\begin{eqnarray}
\mathcal{H}&\equiv& R-(A^{2}_{a}+2A_{b}^{2})+\frac{2}{3}K^{2}-16\pi \rho=0,\label{hamiltonian}\\
\mathcal{M}^{r}&\equiv&\partial_{r}A_{a}-\frac{2}{3}\partial_{r}K+6A_{a}\partial_{r}\chi\nonumber\\
&+&(A_{a}-A_{b})\biggl(\frac{2}{r}+\frac{\partial_{r}b}{b}\biggl)-8\pi S_{r}=0.\label{momentum}
\end{eqnarray}

\subsubsection{Gauge conditions}

In addition to the BSSN spacetime variables, there are two more variables left 
undetermined, the lapse function $\alpha$, and the shift vector $\beta^{i}$. Our code 
can handle arbitrary gauge conditions and for the simulations reported in this paper we 
use the so called {``non-advective 1+log''} condition~\cite{Bona:1997prd} for the 
lapse, and a variation of the {``Gamma-driver''} condition for the shift vector \cite{Alcubierre:2003ab,Alcubierre:2010is}. 

The form of this slicing condition is expressed as
\begin{equation}
\label{eq:1+log1}  
	\partial_{t}\alpha = -2 \alpha K \ . 
\end{equation}
For the radial component of the shift vector, we choose the Gamma-driver 
condition, which is written as 
\begin{align}
\label{eq:shift1}  
	\partial_{t}B^{r} & = \frac{3}{4}\partial_{t}\hat{\Delta}^{r} \ ,\\
\label{eq:shift2}  
 \partial_{t}\beta^{r} & = B^r \ ,
\end{align}
where the auxiliary variable $B^r$ is introduced.

\subsection{Klein-Gordon equation}

In order to solve the Klein-Gordon equation we use two first order variables defined as:
\begin{eqnarray}
\Pi &:=& n^{\alpha}\partial_{\alpha}\Phi=\frac{1}{\alpha}(\partial_{t}\Phi-\beta^{r}\partial_{r}\Phi) \ ,\\
\Psi&:=&\partial_{r}\Phi \ .
\end{eqnarray}
Therefore, using equation (\ref{eq:KG}) we obtain the following system of first order equations: 
\begin{eqnarray}
\partial_{t}\Phi&=&\beta^{r}\partial_{r}\Phi+\alpha\Pi \ ,\\
\partial_{t}\Psi&=&\beta^{r}\partial_{r}\Psi+\Psi\partial_{r}\beta^{r}+\partial_{r}(\alpha\Pi) \ ,\\
\partial_{t}\Pi&=&\beta^{r}\partial_{r}\Pi+\frac{\alpha}{ae^{4\chi}}[\partial_{r}\Psi\nonumber\\
&+&\Psi\biggl(\frac{2}{r}-\frac{\partial_{r}a}{2a}+\frac{\partial{r}b}{b}+2\partial_{r}\chi\biggl)\biggl]\nonumber\\
&+&\frac{\Psi}{ae^{4\chi}}\partial_{r}\alpha+\alpha K\Pi - \alpha \mu^{2}\Phi \ .
\label{eq:sist-KG}
\end{eqnarray}

\subsection{Initial data: solving the constraints}

It has been shown in \cite{Okawa:2014nda} that choosing the maximal slicing condition and a
vanishing scalar field $\Phi$, the momentum constraint is satisfied trivially and the Hamiltonian
constraint can be solved analytically. In order to make the process more dynamic, and allow the black 
hole mass to grow rapidly, we choose instead as initial data a Gaussian distribution for the scalar 
field of the form 
\begin{equation}\label{eq:pulse}
 \Phi=A_0e^{(r-r_0)^2/\lambda^2} \ ,
\end{equation}
with $A_0$ the initial amplitude, $r_0$ the center of the Gaussian and
$\lambda$ its width. The auxiliary first order quantities are
initialized as follows
\begin{eqnarray}
 \Pi(t=0,r)&=&0 \ , \\ 
 \Psi(t=0,r) &=& 2\frac{(r-r_0)}{\lambda^2}A_0e^{(r-r_0)^2/\lambda^2} \ .
 \label{eq:iderivatives}
\end{eqnarray}

We choose a conformally flat metric with $a=b=1$ together with a time symmetry condition $K_{ij}=0$.
With these choices and $\Pi(t=0,r) = 0$ the momentum constraint is satisfied trivially and the Hamiltonian 
constraint~\eqref{hamiltonian} gives an equation for the conformal factor $\psi=e^{\chi}$,
\begin{equation}
 \partial_{rr}\psi + \frac{2}{r}\partial_{r}\psi +2\pi\psi^5\rho=0 \ .
\end{equation}
In order to solve this equation for a back hole, we assume that the conformal factor can be written in a puncture-like form as 
\begin{equation}
 \psi = 1+\frac{M}{2r}+ u(r) \ , 
\end{equation}
and we then substitute this form in the Hamiltonian constraint \eqref{hamiltonian} to obtain 
\begin{equation}
 \partial_{rr}u(r) +\frac{2}{r}\partial_{r}u(r)+2\pi\psi^5\rho = 0 \ . 
\end{equation}
Given a distribution of the scalar field density $\rho$, we solve this ordinary differential equation for $u$ using a 
standard four order Runge-Kutta integrator, assuming that $u\rightarrow 0$ as
$r\rightarrow \infty$ and regularity at the origin. 

\section{Time integration}
\label{sec:NumImpl}

We employ the second-order PIRK method developed by \cite{Isabel:2012arx} to integrate the
evolution equations in time (from time $t^n$ to $t^{n+1}$) and to handle the singular terms that
appear in the evolution equations due to our choice of curvilinear coordinates. Writing a system of PDEs as follows  
\begin{System}
u_t = \mathcal{L}_1 (u, v) \ , \\
v_t = \mathcal{L}_2 (u) + \mathcal{L}_3 (u, v) \ ,
\label{e:system}
\end{System}
where $\mathcal{L}_1$, $\mathcal{L}_2$ and $\mathcal{L}_3$ represent general non-linear 
differential operators, the second-order PIRK method takes the following form:
\begin{System}
	u^{(1)} = u^n + \Delta t \, L_1 (u^n, v^n)º \ ,  \\
	v^{(1)} = v^n + \Delta t \left[\frac{1}{2} L_2(u^n) +
          \frac{1}{2} L_2(u^{(1)}) + L_3(u^n, v^n) \right] \ ,
\end{System}
\begin{System}
	u^{n+1}  = \frac{1}{2} \left[ u^n + u^{(1)} 
+ \Delta t \, L_1 (u^{(1)}, v^{(1)}) \right] \ , \\
	v^{n+1}  =   v^n + \frac{\Delta t}{2} \left[ 
L_2(u^n) + L_2(u^{n+1}) \right.  \\ 
\left. \hspace{2.5cm} +  L_3(u^n, v^n) + L_3 (u^{(1)}, v^{(1)}) \right] \ ,
\end{System}
where we denote by $L_1$, $L_2$ and $L_3$ the corresponding discrete
operators. In particular, we note that $L_1$ and $L_3$ will be treated in an explicit way, 
whereas the $L_2$ operator will contain the singular terms  appearing in the sources of 
the equations and, therefore, will be treated partially implicitly.

This scheme is applied to the Klein-Gordon equation and to the BSSN evolution equations. We 
include all the problematic terms appearing in the sources of the equations, that is stiff source terms that can lead to the development of numerical instabilities (e.g., $1/r$ factors due to spherical coordinates close to $r=0$ even when regular data is evolved), in 
the $L_2$ operator. Firstly, the conformal metric components $a$ and $b$, the quantity 
$X$ (function of the conformal factor), the lapse function $\alpha$, the radial component 
of the shift $\beta^r$, and the scalar field $\Phi$, are evolved explicitly, i.e.~as the generic variable $u$ is 
evolved in the previous PIRK scheme. Secondly, the traceless part of  the extrinsic 
curvature $A_a$, and the trace of the extrinsic curvature $K$, are evolved partially 
implicitly, using updated values of $\alpha$, $a$ and $b$. Then, the quantity $\hat{\Delta}^{r}$, 
and the auxiliary first order quantities of the scalar field, $\Psi$ and $\Pi$, are evolved 
partially implicitly, using the updated values of $\alpha$, $a$, $b$, $\beta^r$, $\psi$, 
$A_a$, $K$, and $\Phi$. Finally, $B^r$ is evolved partially implicitly, using the 
updated values of $\hat{\Delta}^{r}$. We note that the matter source terms are always included 
in the explicitly treated parts. In the Appendix~\ref{appendix}, we give the exact 
form of the source terms included in each operator.

\begin{figure}
\begin{minipage}{1\linewidth}
\includegraphics[width=1.11\textwidth, height=0.3\textheight]{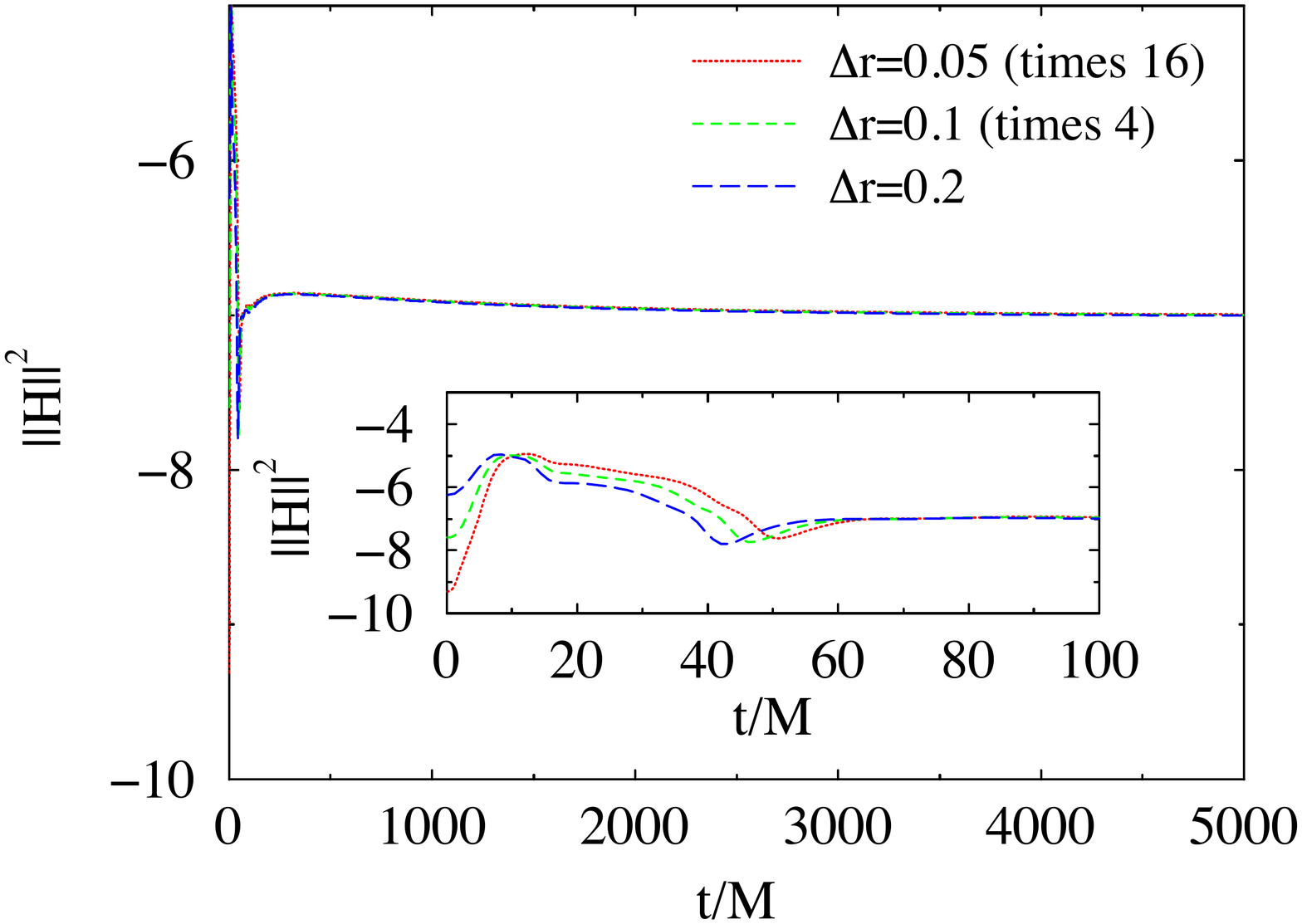} 
\caption{Time evolution of the L2 norm of the Hamiltonian constraint for an initial scalar field pulse with 
$A_0=0.1$ and mass $M\mu=0.1$. The plot shows results for three different resolutions, 
rescaled by the factors corresponding to second-order convergence. The inset shows a magnified 
view of the initial 100M in the evolution. At $t=0$ the convergence is to fourth order.}
\label{fg:Converg}
\end{minipage}
\end{figure}

\begin{table*}
\caption{Initial parameters for the scalar field around the black hole with initial bare mass $M=1$. 
The initial Gaussian is located at $r_0=10M$ with half-width $\lambda=5$.
In all cases the extraction radius is $r_{ext}=100M$. $M_{\rm{AH}}$
corresponds to the final mass of the apparent horizon of the black
hole, $M_{\rm{ADM}}$ is the initial ADM mass of the system and $E_{0}$ is the initial energy of the scalar field. }
\label{tab:zzz}
\begin{ruledtabular}
\begin{tabular}{ccccccccc}
Model&$M\mu$&$A_0$&
\multicolumn{3}{c}{$M\omega$}&$M_{\rm{AH}}$&$M_{\rm{ADM}}$&$E_{0}$\\
\cline{4-6}
& & &1&2 &3\\

\hline
1&0.08&0.1&0.09608&0.09772&-&2.03&3.23&2.28\\
2&0.09&0.1&0.10876&0.11063&-&2.07&3.31&2.37\\
3&0.10&0.0005&0.09946&0.09985&-&1.0&1.000062&6.4E-6\\
4&0.10&0.005&0.09948&0.09989&-&1.0&1.0062&6.4E-6\\
5&0.10&0.05&0.09984&0.10455&0.10535&1.20&1.614&0.64\\
6&0.10&0.1&0.12117&0.12412&-&2.18&3.40&2.46\\
7&0.11&0.1&0.13382&-&-&2.23&3.50&2.56\\
8&0.12&0.1&0.15253&-&-&2.36&3.61&2.66\\
9&0.13&0.1&0.16734&-&-&2.48&3.73&2.79\\
10&0.14&0.1&0.18266&-&-&2.55&3.85&2.91\\
11&0.15&0.0005&0.14797&0.14951&0.14981&1.0&1.00008&8.2E-5\\
12&0.15&0.005&0.14806&0.14957&0.14981&1.0&1.008&8.2E-3\\
13&0.15&0.05&0.15062&0.15983&0.16043&1.37&1.77&0.8\\
14&0.15&0.1&0.19865&-&-&2.68&3.98&3.04\\
15&0.16&0.1&0.21529&-&-&2.82&4.13&3.19\\
16&0.17&0.1&0.23257&-&-&2.96&4.27&3.34\\
17&0.18&0.1&0.25079&-&-&3.01&4.43&3.49\\
18&0.19&0.1&0.26980&-&-&3.25&4.59&3.65\\
19&0.20&0.0005&0.19508&0.19883&0.19949&1.0&1.0001&1.1E-4\\
20&0.20&0.005&0.19513&0.19902&0.19973&1.0&1.010&1.1E-2\\
21&0.20&0.05&0.20147&0.21867&-&1.57&1.99&1.03\\
22&0.20&0.1&0.28959&-&-&3.46&4.76&3.82\\
23&0.30&0.0005&0.29812&-&-&1.0&1.00017&1.8E-4\\
24&0.30&0.005&0.29843&-&-&1.0&1.017&1.8E-2\\
25&0.30&0.05&0.34413&-&-&2.14&2.61&1.67\\
\end{tabular}
\end{ruledtabular}
\end{table*}

\begin{figure}
\begin{center}
 \includegraphics[width=0.52\textwidth]{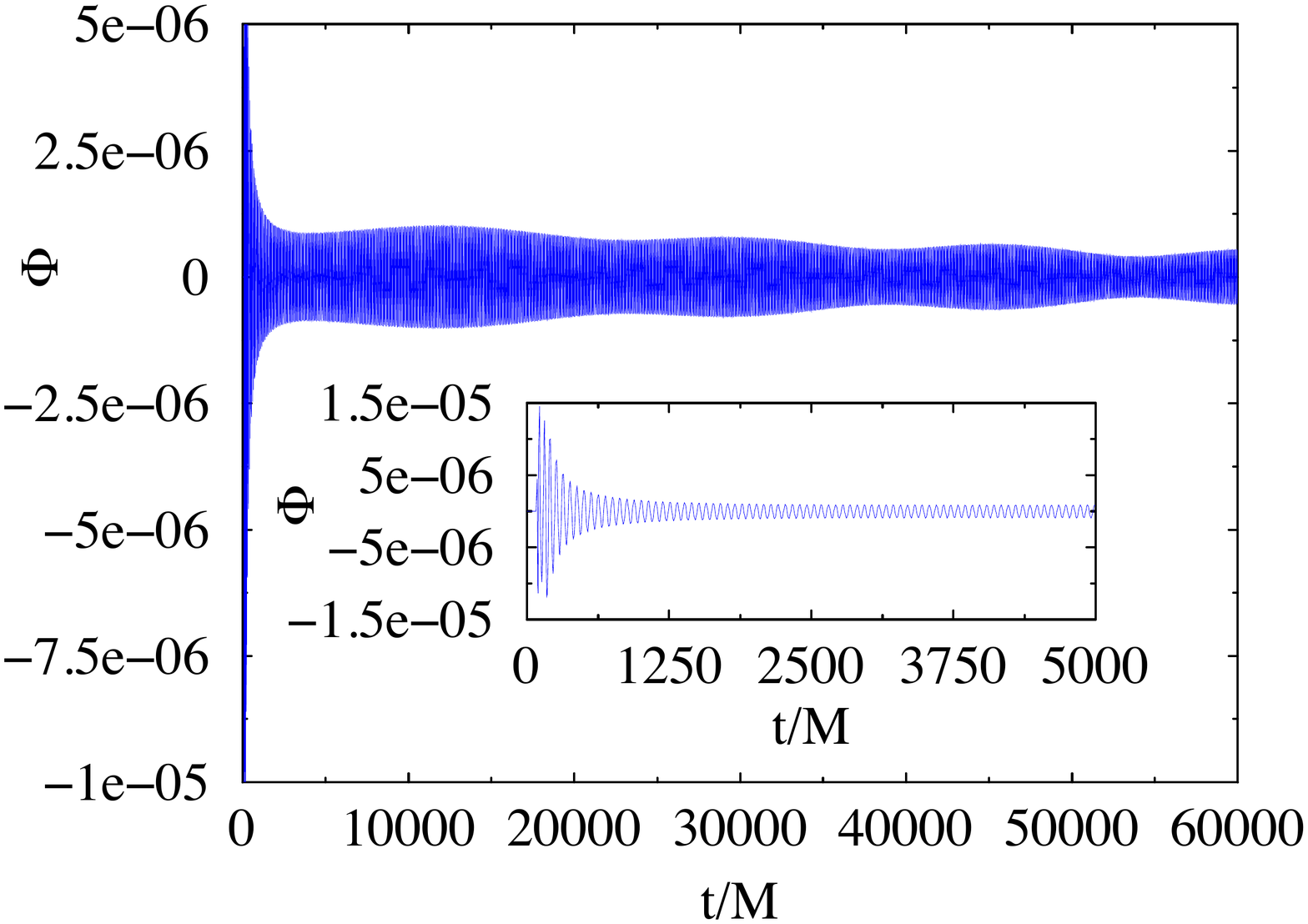}
 \vskip -1.3cm
 \includegraphics[width=0.52\textwidth]{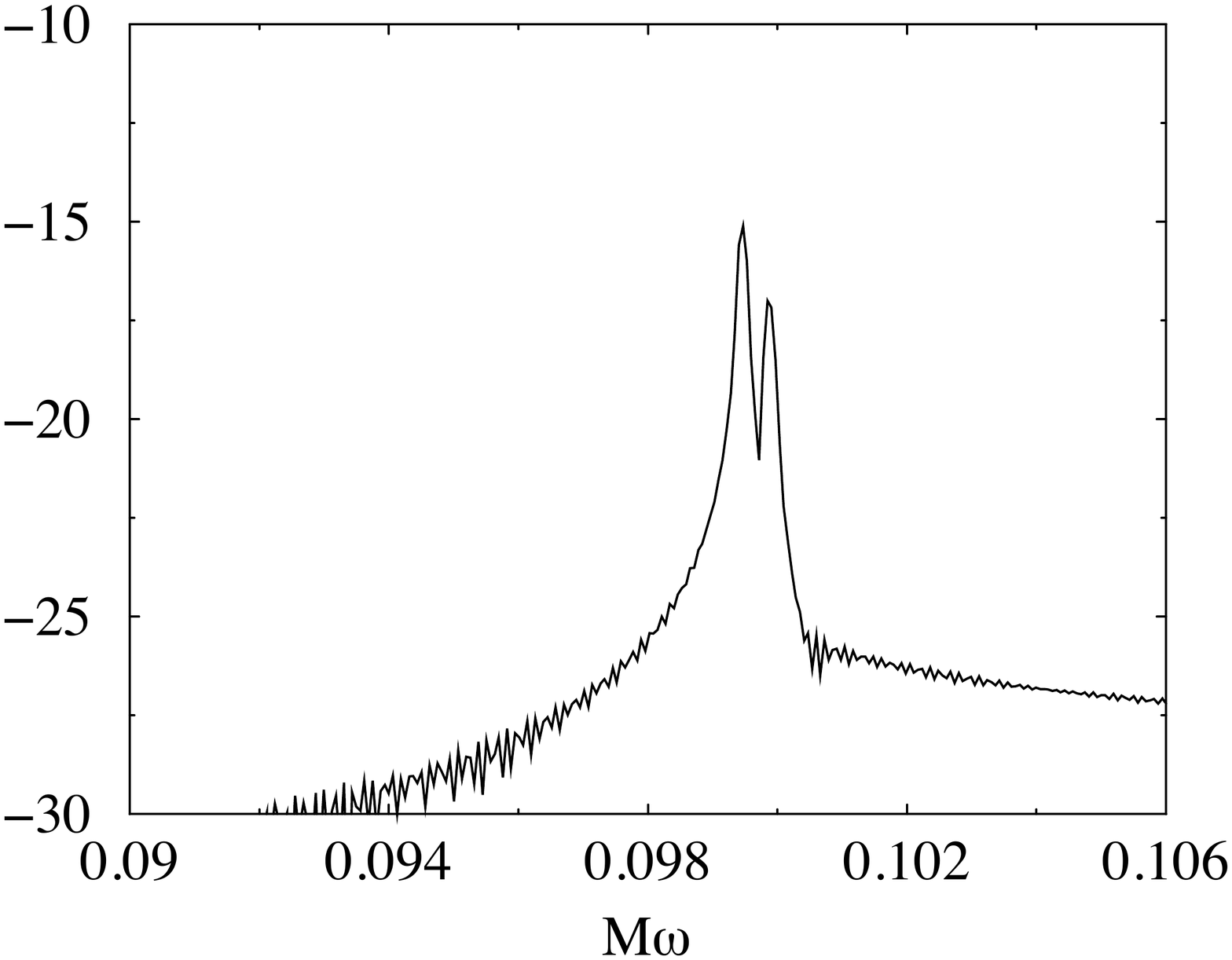}
\caption{{\it Upper panel}: Time evolution of the scalar field with mass $M\mu=0.1$ and $A_{0}=0.0005$ corresponding to the model 3 for an observer at $r_{\rm{ext}}=100M$. The inset shows a magnified view of the initial 5000M in the evolution. {\it Lower panel}: Fourier transform of the evolution of the scalar field. The units in the vertical axis are arbitrary.}
\label{fg:SF3}
\end{center}
\end{figure}

\begin{figure}[t]
\begin{center}
 \includegraphics[width=0.52\textwidth]{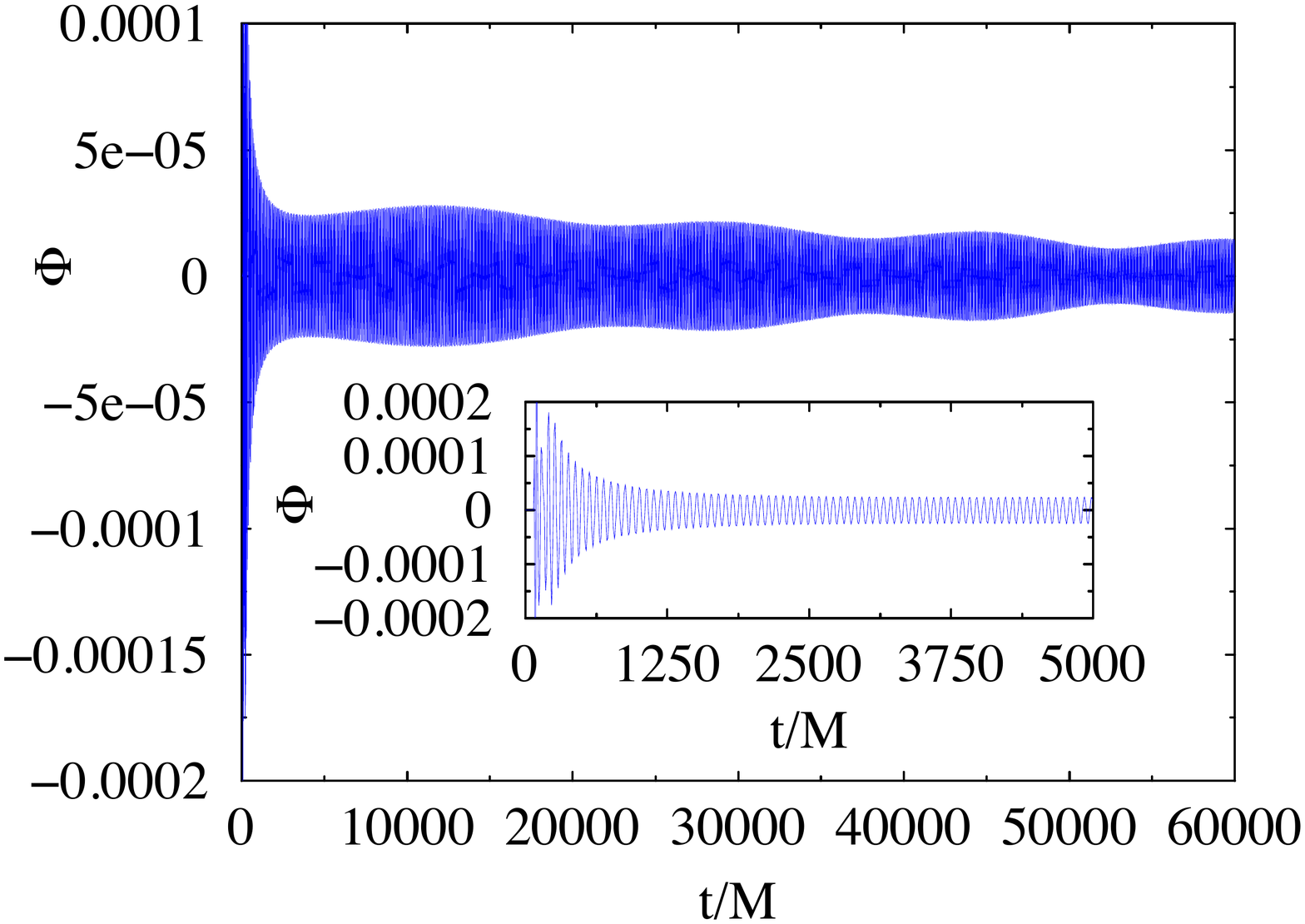}
 \vskip -1.3cm
 \includegraphics[width=0.52\textwidth]{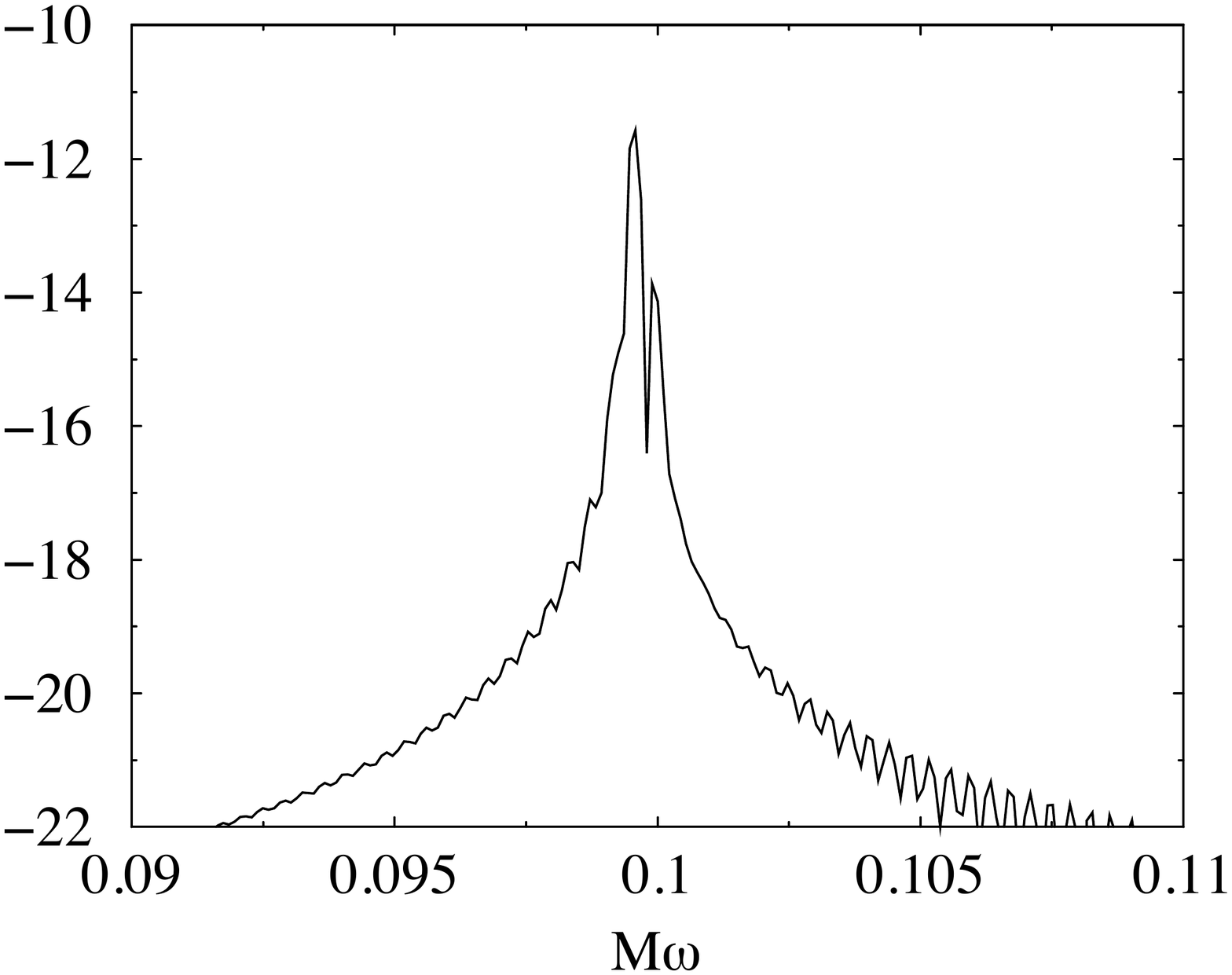}
\caption{Same as Fig.~\ref{fg:SF3} but for model 4.}
\label{fg:SF4}
\end{center}
\end{figure}

\begin{figure}[t]
\begin{center}
 \includegraphics[width=0.52\textwidth]{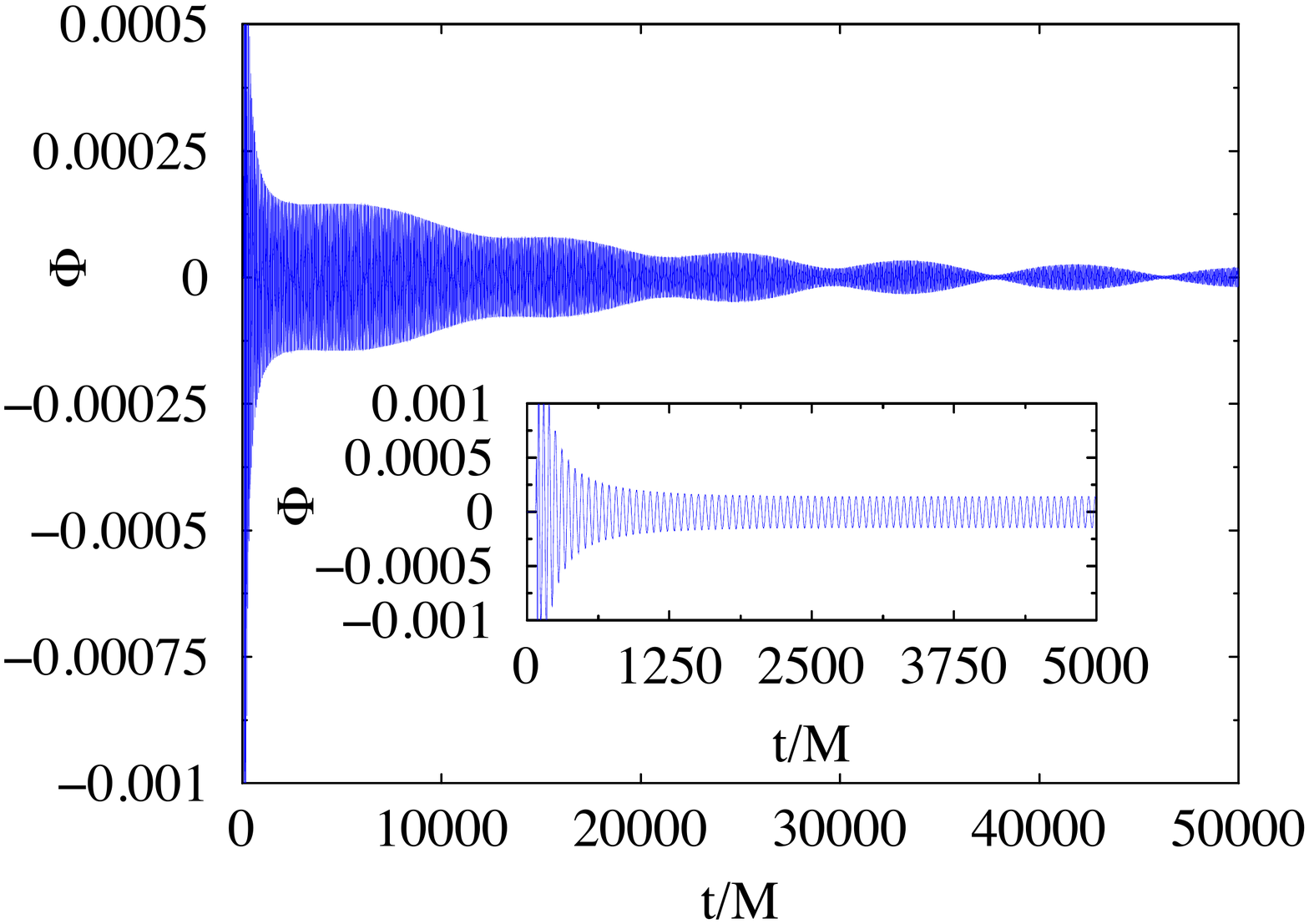}
 \vskip -1.3cm
 \includegraphics[width=0.52\textwidth]{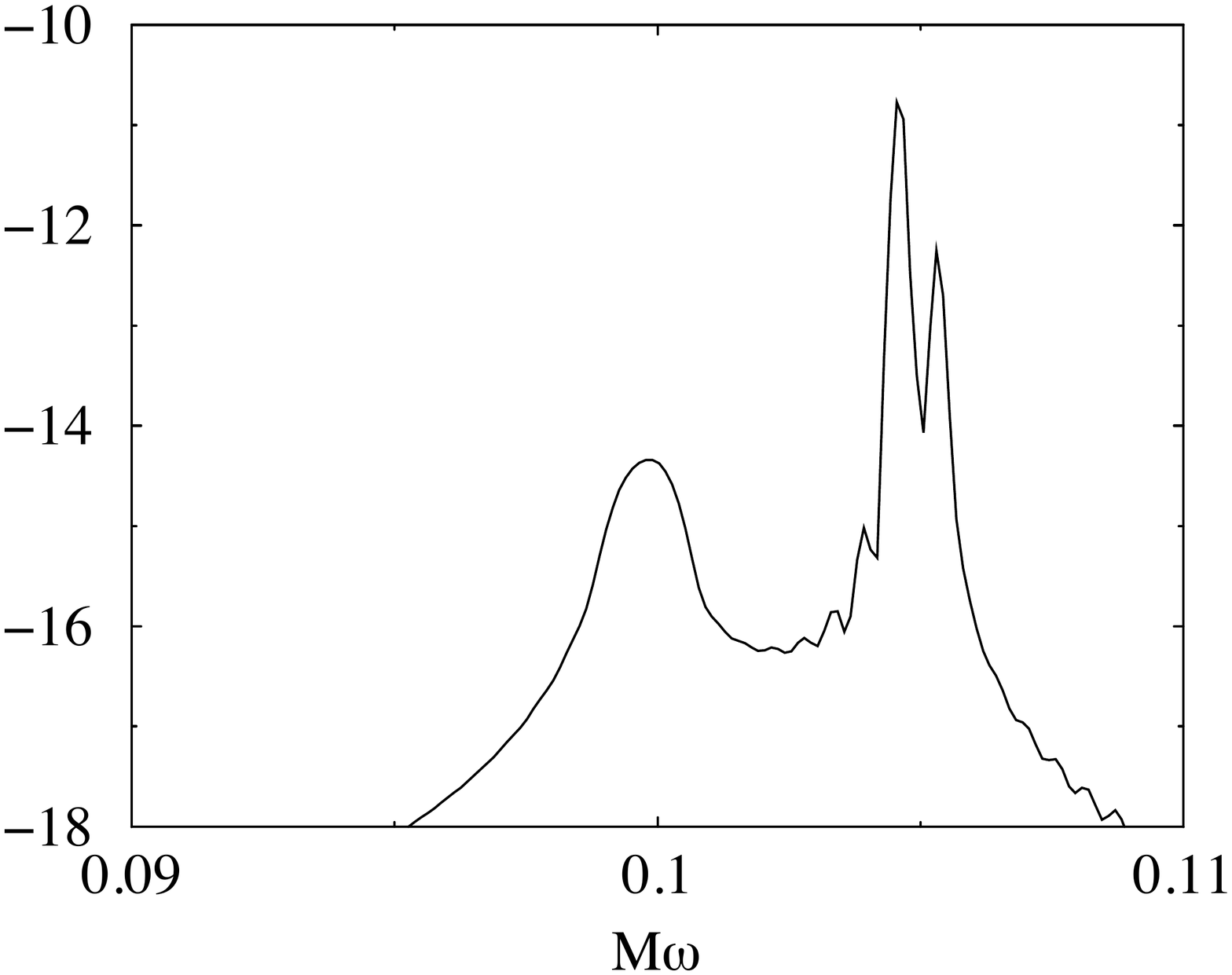}
\caption{Same as Fig.~\ref{fg:SF3} but for model 5.}
\label{fg:SF5}
\end{center}
\end{figure}

\section{Numerical Results} \label{sec:num_results}

\subsection{Initial models and convergence}

We set up puncture-like initial data representing a stationary black hole with a surrounding scalar field 
cloud. Whereas part of the initial scalar field is absorbed by the black hole, another part lingers for a long 
time in the form of a long-lived quasi-bound state which continuously falls into the black hole. 
We have set up and evolved 25 different models exploring different values of the initial amplitude of the pulse 
$A_{0}=\lbrace0.0005,0.005,0.05,0.1\rbrace$ and different values of the scalar field mass $M\mu=\lbrace0.08:0.3\rbrace$ 
to study their influence in the evolution. Regarding the initial pulse, all models are placed at the same initial location 
$r_{0}=10M$ and share the same width
$\lambda=5$. The values for the amplitude are chosen so that the self-gravitation of the scalar field becomes important, 
far away from the test field regime. The bare mass of the black hole is initially set to $M=1$.

The grid resolution is set to $\Delta r=0.2M$ and the step time is given by $\Delta t=0.5\Delta r$ or 
$\Delta t=0.3\Delta r$. The second choice is needed to obtain long-term stable simulations for the models with 
larger amplitude. The final time of the numerical evolutions varies between $2\times10^4M$ 
for the large amplitude models and about $6\times10^4M$ for the small amplitude ones. The outer boundary of the
computational domain is placed at $r=6\times10^4M$, far enough so 
it does not affect the dynamics in the inner region.

In Table~\ref{tab:zzz} we present the different choices for the initial parameters of 
the models we have evolved, that is, the mass of the scalar field particle $M\mu$ and the amplitude of the pulse $A_0$. 
The frequencies $M\omega$ of the dominant modes are labeled according to their strength on the power spectrum.
The table also contains the mass of the apparent horizon (AH) at the end of the simulation $M_{\rm{AH}}$, defined as 
$M_{\rm{AH}}=\sqrt{\mathcal{A}/16\pi}$, where $\mathcal{A}$ is the area of the AH, the Arnowitt-Deser-Misner (ADM) mass 
$M_{\rm{ADM}}$ of the whole system
\begin{equation}\label{eq:admmass}
M_{\rm{ADM}}=-\frac{1}{2\pi}\lim_{r\rightarrow \infty}\oint_{S}\partial_{j}\psi dS^{j}\quad,
\end{equation}
and the initial energy of the scalar field $E_0$
\begin{equation}\label{eq:scalar}
E=\int_{2M}^{\infty}\rho dV  \ ,
\end{equation}
where $\rho$ is defined in Eq.~(\ref{eq:rho}). We have checked that the total ADM mass of the system $M_{\rm{ADM}}$ computed at the initial time slice using Eq.~(\ref{eq:admmass}) is equal to the sum of the initial black hole mass plus the initial energy of the scalar field Eq.~(\ref{eq:scalar}). The discrepancy is at most 2\% of the ADM mass.

In order to test the convergence of the code we performed three simulations with different resolutions $\Delta r = \lbrace0.2M,0.1M,0.05M\rbrace$. In Fig.~\ref{fg:Converg} we plot the rescaled evolution of the L2 norm of the Hamiltonian 
constraint for the particular choice of the initial amplitude $A_0=0.1$ and the scalar field mass $M\mu=0.1$, obtaining 
the expected second-order convergence of our PIRK time-evolution scheme. The same result is achieved irrespective of the model.

\begin{figure}[t]
\begin{center}
 \includegraphics[width=0.52\textwidth]{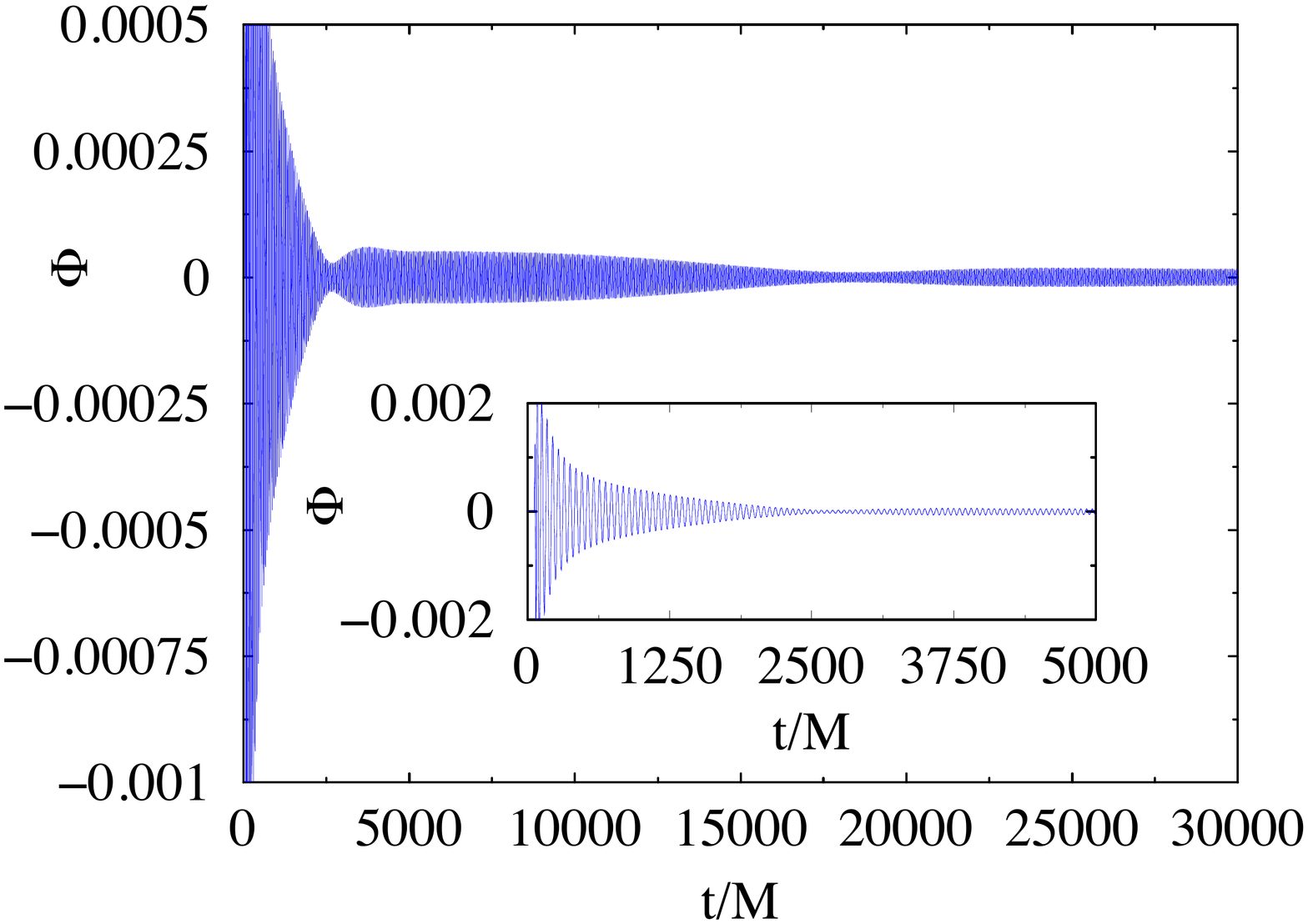}
 \vskip -1.3cm
 \includegraphics[width=0.52\textwidth]{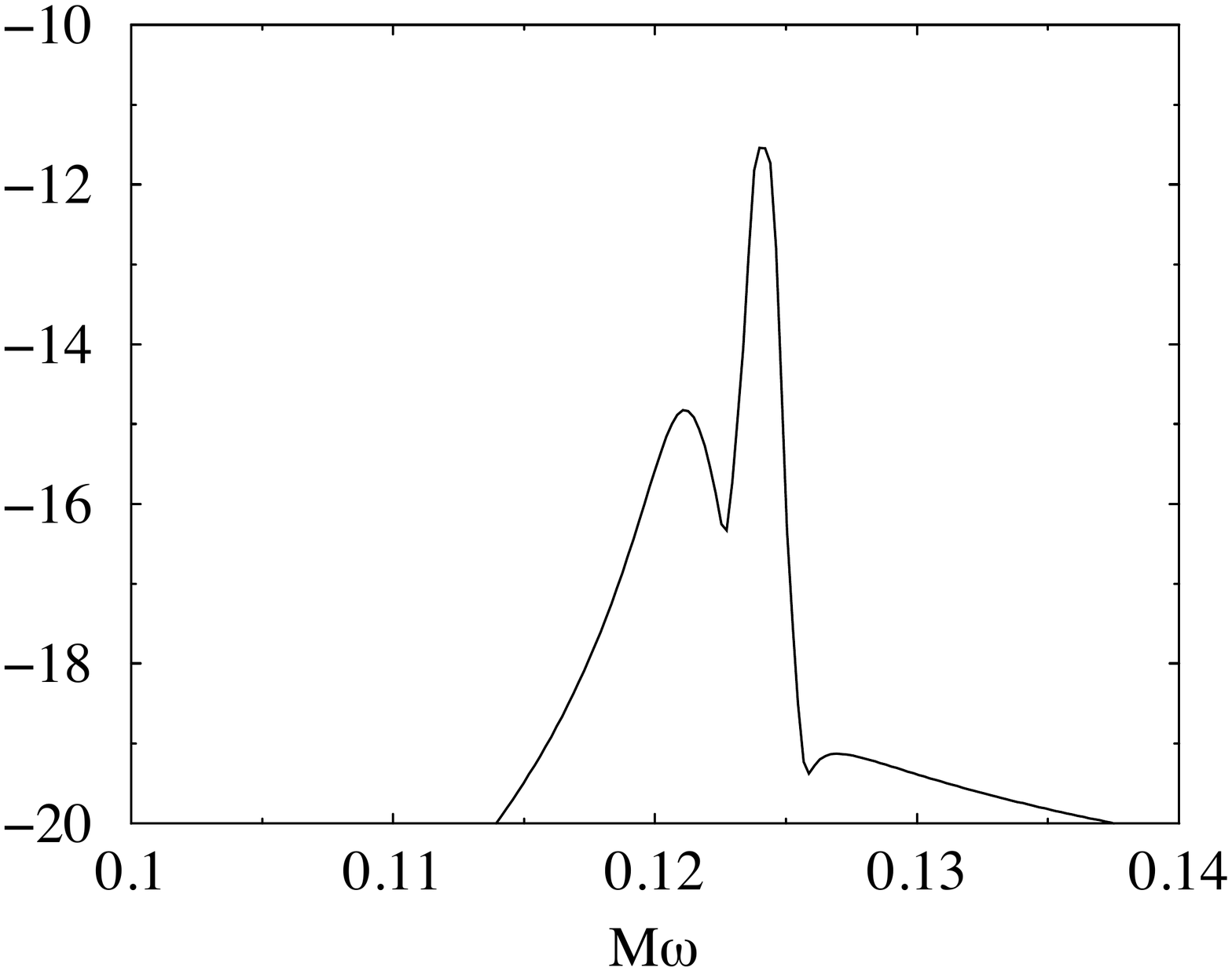}
\caption{Same as Fig.~\ref{fg:SF3} but for model 6.}
\label{fg:SF6}
\end{center}
\end{figure}

\subsection{Non-linear quasi-bound states}

We evolve the Einstein-Klein-Gordon system using the initial data given by Eqs.~\eqref{eq:pulse}-\eqref{eq:iderivatives}. 
In order to analyze the results of the simulations we extract a time series for
the scalar field amplitude at an observation point with a fixed radius $r_{\rm{ext}}$. Thus, 
to identify the frequencies at which the field oscillates we perform a Fast Fourier
transform after a given number of time steps and obtain the power spectrum.   
For some models we were able to obtain not only the fundamental frequency of oscillation but also some of the overtones.

It is well known that the frequency resolution
is inversely proportional to the simulation time and hence longer runs
yield to more accurate frequencies. Thanks to the computational advantage provided by spherical symmetry and to the use of the PIRK method
we were able to compute the frequencies at which the scalar field oscillates, confirming the existence 
of long-lived spherical states in our nonlinear evolutions even for large initial amplitudes of the scalar field. 

The power spectrum obtained from the Fourier transform shows a set of distinct frequencies. In order to validate our results 
we contrast the numerical values of the frequency of the small amplitude configurations with the 
ones obtained using Leaver's continued fraction technique \cite{Leaver:1985ax} in the frequency 
domain in the test field approximation. 

In figures \ref{fg:SF3} to \ref{fg:SF6} we plot the evolution of the scalar field seen by an observer 
at $r_{\rm{ext}}=100M$ for the models 3-6 of our sample along with their corresponding Fourier transforms. 
In all of the plots the field is seen to be clearly 
oscillating. Moreover, all of these models show a distinctive beating pattern due to the presence of overtones, as described in \cite{Dolan:2012yt,Witek:2012tr,Zhang:2013ksa,Degollado:2013bha} 
for the case of fixed background computations and in \cite{Okawa:2014nda} in the non-linear regime. Within the computational times we can afford, such beatings and overtones cannot be resolved only in models 7-10, 14-18, and 22-25.
For the small amplitude models 3-12, 14, 19, 20 and 23 the frequencies are very close to 
the values obtained in the test field regime. As shown in Table~\ref{tab:zzz} the 
frequency grows as the amplitude is increased.

\begin{figure}
\begin{center}
 \includegraphics[width=0.5\textwidth]{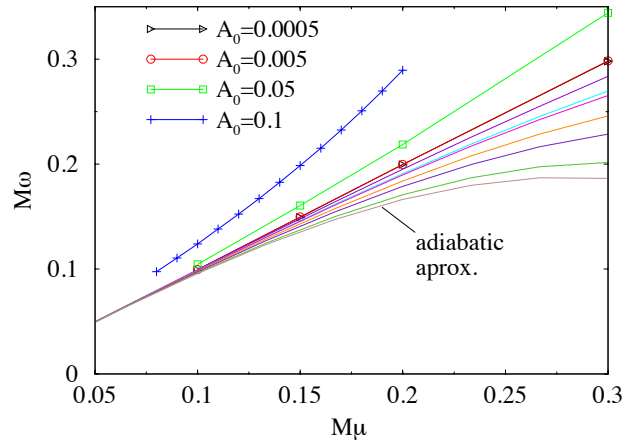}\vspace{-0.5cm}
\caption{Frequency $\omega$ as a function of the scalar field mass $\mu$ for the models shown in 
Table~\ref{tab:zzz}. The semi-analytic relation \eqref{eq:furuhashi} for several values of $M$ is also plotted. 
For models with small initial $A_0$ the numerical frequencies match the values of the semi analytical values 
showing consistency with the test field approximation and the limit $M\mu\ll1$.}
\label{fg:freq_vs_mu}
\end{center}
\end{figure}

By using a matching technique, Furuhashi and Nambu showed analytically in 
\cite{Furuhashi:2004jk} that in the limit $M\mu\ll1$ the real part of the frequency of quasi-bound states depends on the 
mass parameter $\mu$ as
\begin{equation}
 \omega\approx \mu\left[1-\frac{1}{2}(M\mu)^2 \right]\ .
\label{eq:furuhashi}
\end{equation}
In Fig.~\ref{fg:freq_vs_mu} we show a plot of the frequencies of all the configurations of Table~\ref{tab:zzz} as a
function of $\mu$.  We also consider in this plot the adiabatic approximation for the frequency given by 
Eq.~\eqref{eq:furuhashi} for different values of the mass of the black hole $M$. For small values of $A_0$, the 
corresponding curve is indistinguishable from the one of the test field approximation. However, for greater values, 
the results from the non-linear approach not only deviate from the quasi-adiabatic approximation but they also 
follow the opposite trend. The frequency is greater than the frequency of the test field limit, whereas the adiabatic 
approach holds that the frequency should be smaller. This discrepancy may be due to the violation of the condition 
$M\mu\ll1$ used in the analytical approximation. Nevertheless, the extrapolation of our results for the scalar field mass $\mu$ in the regime compatible with scalar field dark matter models, $M\mu\sim10^{-6}$, supports the validity of the model.

\subsection{Spacetime characterization}

\begin{figure*}[ht]
\begin{center}
\includegraphics[width=0.55\textwidth, height=0.35\textheight]{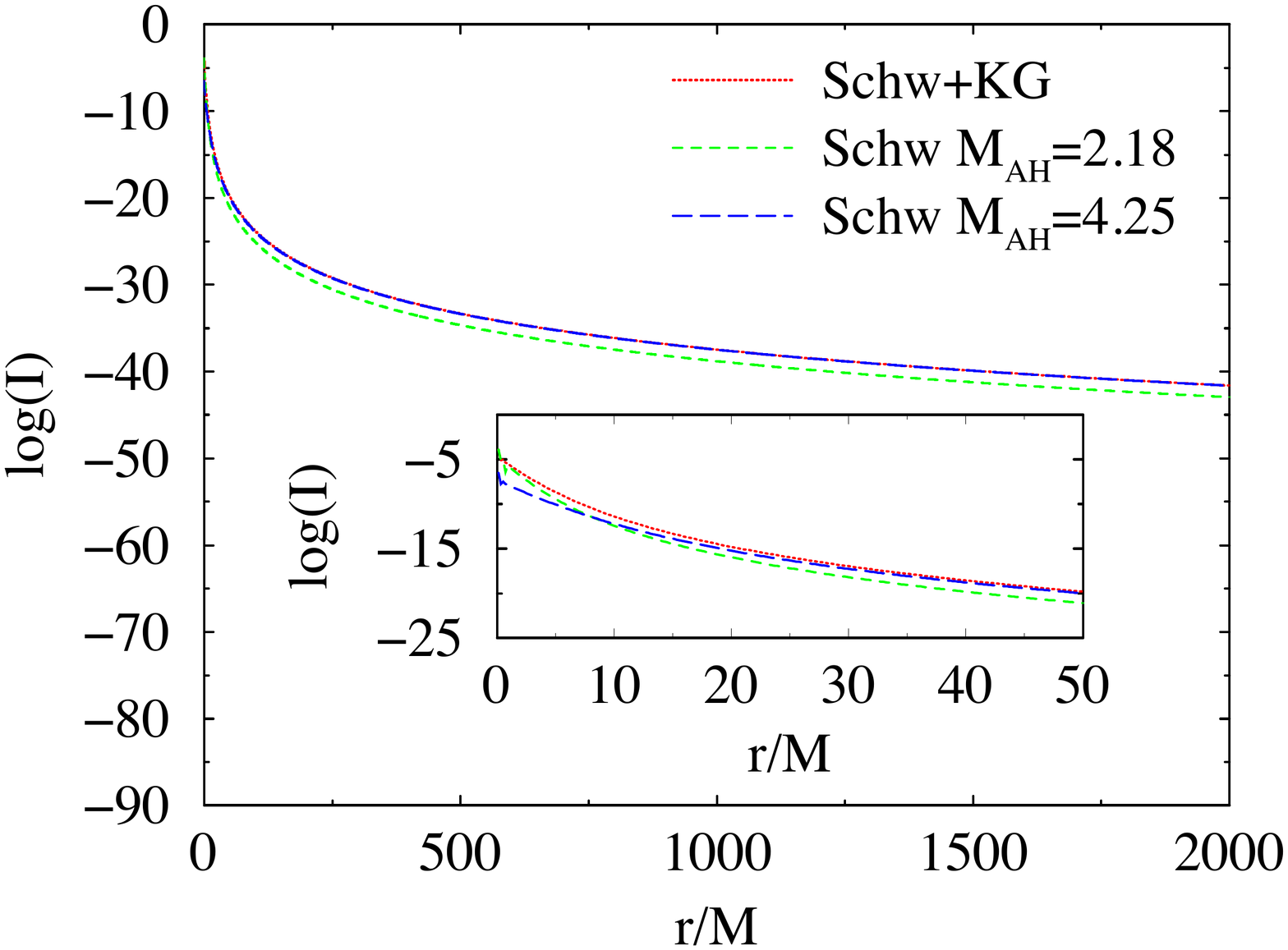}
\hspace{-1.89cm}\includegraphics[width=0.55\textwidth, height=0.35\textheight]{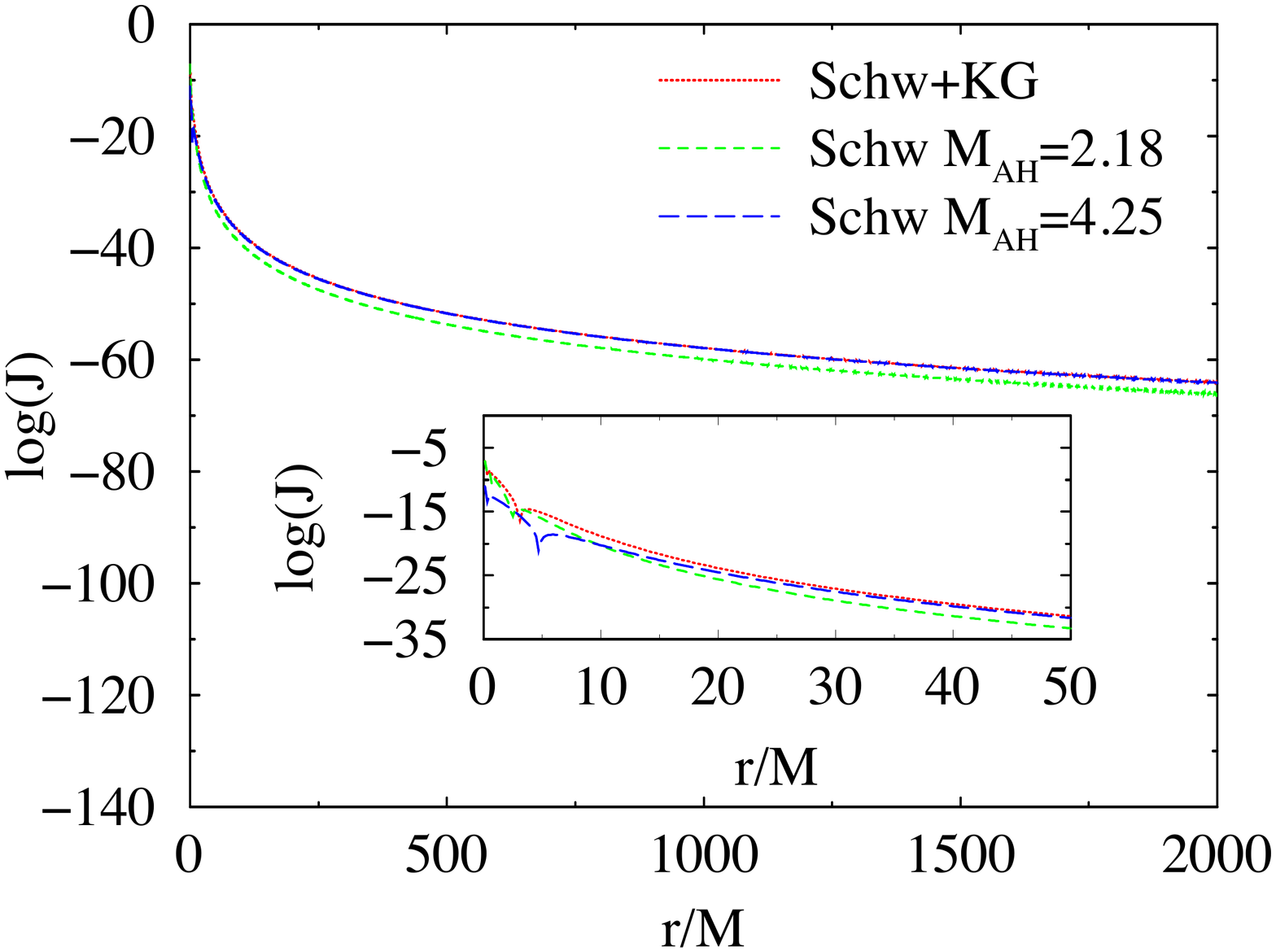}
\caption{Radial profiles of the Weyl invariants $I$ (left) and $J$ (right) for model 6 in Table~\ref{tab:zzz} at $t=9000M$. 
The red curve corresponds to our numerically evolved spacetime while the green and blue curves correspond to 
analytic (static) Schwarzschild spacetimes of mass $M_{\rm{AH}}=2.18$ and $4.25$, respectively. The inset shows 
a region  close to the black hole horizon where the numerical solution differs considerably from a stationary 
Schwarzschild black hole.}
\label{fig:m_6_IJ} 
\end{center}
\end{figure*}

Spacetime invariants are useful quantities to compare numerical solutions with known exact 
solutions, see for instance Ref.~\cite{Baker:2000zm} for an instructive way to use them during the  
formation of a single black hole. With this aim we have used the invariants $I$ and $J$ defined 
as~\cite{Alcubierre08a}
\begin{eqnarray}
I&=& \frac{1}{2}\left[ (E_{ij}E^{ij} -B_{ij}B^{ij} )-2iE_{ij}B^{ij}\right] \, , \\ \nonumber
J&=&-\frac{1}{6}\left[  ( E^{i}{}_{k}E^{jk}-3B^{i}{}_{k}B^{jk} )E_{ij}\right.\\
 &&\left.+i(B^{i}_{k}B^{jk}-3E^{i}_{k}E^{jk})B_{ij}    \right]\, ,  \nonumber
\end{eqnarray}
to characterize our numerical spacetime. In these expressions $E_{ij}$ and $B_{ij}$ are the electric and magnetic parts of the Weyl tensor. We have computed these two quantities to use them as a measure of the deviation of our numerical solution from 
that corresponding to a Schwarzschild black hole. As expected, the numerical spacetime for the cases with a scalar field pulse 
of sufficiently small amplitude resembles the Schwarzschild solution once most of the scalar field has accreted on the black
hole. The situation changes in a significant way when the amplitude increases. In the two plots of Fig.~\ref{fig:m_6_IJ} the red solid curves correspond to the invariants $I$ 
(left panel) and $J$ (right panel) for our model 6 at $t=9000M$. These radial profiles are compared with the corresponding purely Schwarzschild 
invariants for a black hole with the same final mass ($M_{\rm{AH}}=2.18$), as indicated by the green dashed curves in the same figure.  
The deviation of our numerical spacetime from an analytic Schwarzschild black hole spacetime with equivalent mass is evident. 
An important remark is that both invariants may however be superimposed, at sufficiently large distances, if we choose a larger Schwarzschild black hole mass. The blue dashed curve in Fig.~\ref{fig:m_6_IJ} shows that that is indeed the case if the mass of
the Schwarzschild black hole is chosen to be $M_{\rm{AH}}=4.25$. The discrepancy is still noticeable near the black hole horizon, as the inset in Fig.~\ref{fig:m_6_IJ} clearly makes visible.

\begin{table}
\caption{Mass of the black hole spacetimes corresponding to the six largest amplitude models of our sample. 
Columns three and four indicate, respectively, the masses of our numerically evolved spacetimes and those 
of the analytic Schwarzschild black holes that would induce the same radial profiles (degeneracy) in the Weyl 
invariants.}\label{tab:masses}
\begin{ruledtabular}
\begin{tabular}{ccccc}
Model&$M\mu$&$M_{\rm{AH1}}$&$M_{\rm{AH2}}$&$M_{\rm{AH1}}/M_{\rm{AH2}}$\\

\hline
6&0.10&2.18&4.25&1.95\\
8&0.12&2.36&5.0&2.12\\
10&0.14&2.55&6&2.35\\
15&0.16&2.82&7.25&2.57\\
17&0.18&3.1&8.75&2.82\\
22&0.20&3.46&11.0&3.18\\
\end{tabular}
\end{ruledtabular}
\end{table}


This difference may have important implications when studying the motion of test particles around black holes. 
In particular, there would be a degeneracy in the characterization of the underlying spacetime generating the 
gravitational attraction of a test particle. One observer at infinity might infer, due to the movement of a test particle, 
that the central object is either one isolated Schwarzschild black hole with mass $M_1$ or a transient (non-static) 
state made up of a scalar cloud around a black hole with mass $M_2$. These two possibilities can be regarded 
as well as two different, very long-lived solutions of Einstein equations that induce the same motion of test particles, 
either the vacuum Schwarzschild solution or that of a dynamical spherically-symmetric black hole surrounded 
by a scalar field.

In Table~\ref{tab:masses} we summarize the results for all the 6 cases of our sample of models with the largest 
initial amplitudes and for which the aforementioned degeneracy is most apparent. The third column corresponds 
to the final numerical black hole mass, $M_{\rm{AH1}}$. The fourth column reports the equivalent Schwarzschild 
mass $M_{\rm{AH2}}$ that would induce the same radial profiles of the Weyl invariants at large distances. The 
second mass is extracted by fitting the invariants at large distances and therefore it has not to be taken as an 
accurate measure but rather as an approximate value. The ratio $M_{\rm{AH1}}/M_{\rm{AH2}}$ grows with 
$M\mu$, as shown in the fifth column of the table. We note that it remains to be seen whether the spacetime 
degeneracy found in our simulations is still important, or even exists in the first place, for values of the scalar 
field mass $\mu$ compatibles with dark matter models, namely $M\mu\sim10^{-6}$. As mentioned in the Introduction 
the timescales needed to computationally disclose the answer cannot be reached even with linear perturbation 
codes. 

\section{Conclusions} \label{sec:conclusions}

In this paper we have solved numerically the Einstein-Klein-Gordon system in spherical symmetry using 
spherical coordinates with a PIRK numerical scheme. Our study has been focused on investigating within 
a fully non-linear setup whether quasi-stationary scalar field configurations may be found in the form of clouds 
(or hairy ``wigs") surrounding dynamical black holes. Such configurations have been recently put forward 
by~\cite{Barranco:2012qs, Barranco:2013rua} as a plausible model to describe dark matter halos in galaxies. 
Through a computational study based on perturbative approaches, these authors found indeed such 
configurations to be long-lived.

In our non-linear study we have performed a large number of (second-order) accurate and long-term stable
simulations of dynamical non-rotating black holes surrounded by self-gravitating scalar fields. The models of 
our sample have been suitably parameterized to span a broad range of cases from the test field regime to the 
fully non-linear regime. Confirming earlier findings from perturbation theory we have found that also in the 
case of highly dynamical spacetimes (i.e.~those consisting of a black hole and a rich scalar field environment) 
there are states which closely resemble the quasi-bound states in the test field approximation 
of~\cite{Barranco:2012qs, Barranco:2013rua}. In addition, we have been able to characterize the resulting 
spacetimes by analyzing the Weyl invariants $I$ and $J$ and have compared them to analytical values of 
Schwarzschild black hole spacetimes. Our results have revealed a degeneracy in plausible long-lived solutions 
of Einstein equations that induce the same motion of test particles, either with or without the existence of 
quasi-bound states. By performing a Fourier transform of the time series of our numerical data we have been 
able to characterize the scalar field states by their distinctive oscillation frequencies. It has been found that the 
scalar field oscillates with different well-defined frequencies, and that its time evolution produces in most of 
our models a characteristic beating pattern due to the non-linear combination of two frequencies with close 
enough values.  Such scalar field oscillations may have important imprints in a number of astrophysical 
scenarios. For instance, the fully three-dimensional dynamical evolutions performed in~\cite{Okawa:2014nda} 
showed that the interaction of the scalar field with the central black hole results in both scalar field and 
gravitational radiation. Recently~\cite{Degollado:2014vsa}, using first-order perturbation theory, found that 
the gravitational response is already present even at the linear level. Since our current study is limited to 
spherical symmetry, we could not perform such analysis in the present work, a task which we defer to a future 
investigation.

\section*{Acknowledgements}

This work has been supported by CONACyT-M\'exico, by the Deutsche Forschungsgemeinschaft (DFG) through its Transregional Center SFB/TR7 ``Gravitational Wave Astronomy'', by the Spanish MINECO (AYA2013-40979-P) 
and by the Generalitat Valenciana (PROMETEOII-2014-069). The computations have been performed at the Servei
d'Inform\`atica de la Universitat de Val\`encia.

\appendix
\section{Source terms}
\label{appendix}

The evolution Eqs.~(\ref{eq:X})-(\ref{eq:b}), (\ref{eq:K})-(\ref{eq:Deltar}), (\ref{eq:1+log1})-(\ref{eq:shift2}) are evolved using a second-order PIRK method, described in Sec.~III. In this Appendix the source terms 
included in the explicit or partially implicit operators are detailed.

Firstly, $a$, $b$, $X$, $\alpha$, $\beta^r$ and $\Phi$, are evolved explicitly, i.e., all the source terms of the evolution equations of these variables are included in the $L_1$ operator of the second-order PIRK method.

Secondly, $A_a$ and $K$, are evolved partially implicitly, using updated values 
of $\alpha$, $a$ and $b$. More precisely, the corresponding $L_2$ and $L_3$ 
operators associated with the evolution equations for $A_a$ and $K$ read:
\begin{align}
	L_{2(A_a)} &= - \left(\nabla^{r}\nabla_{r}\alpha 
- \frac{1}{3}\nabla^{2}\alpha\right) 
+ \alpha\left(R^{r}_{r} - \frac{1}{3}R\right) \ , \\
	L_{3(A_a)} &= \beta^{r}\partial_{r}A_{a} + \alpha K A_{a} 
- 16\pi\alpha(S_a - S_b) \ , \\
	L_{2(K)} &= - \nabla^{2}\alpha \ , \\
	L_{3(K)} &= \beta^{r} \partial_{r}K  
+ \alpha(A_{a}^{2} + 2A_{b}^{2} + \frac{1}{3}K^{2}) \nonumber \\
& + 4\pi\alpha(\rho + S_{a} + 2S_{b}) \ .
\end{align}
Next, $\hat{\Delta}^{r}$, $\Psi$ and $\Pi$ are evolved partially implicitly, using the updated values 
of $\alpha$, $a$, $b$, $\beta^r$, $\psi$, $A_a$, $K$ and $\Phi$.
Specifically, the corresponding $L_2$ and $L_3$ operators associated with the 
evolution equation for $\hat{\Delta}^{r}$, $\Psi$ and $\Pi$ are given by:
\begin{align}
	L_{2(\hat{\Delta}^{r})} &= \frac{1}{a}\partial^{2}_{r}\beta^{r} 
+ \frac{2}{b}\partial_{r}\left(\frac{\beta^r}{r}\right)
+ \frac{\sigma}{3 a}\partial_{r}(\hat{\nabla}_m\beta^{m}) \nonumber \\
  & - \frac{2}{a}(A_{a}\partial_{r}\alpha + \alpha\partial_{r}A_{a}) 
- \frac{4\alpha}{r b}(A_{a}-A_{b}) \nonumber \\
	& + \frac{\xi \alpha}{a} \left[\partial_{r}A_{a} 
- \frac{2}{3}\partial_{r}K + 6A_{a}\partial_{r}\chi  \right. \nonumber \\
  & \left. + (A_{a}-A_{b})\left(\frac{2}{r} + \frac{\partial_{r}b}{b}\right)
\right] \ , \\
	L_{3(\hat{\Delta}^{r})} &= \beta^{r}\partial_{r}\hat{\Delta}^{r}
- \hat{\Delta}^{r}\partial_{r}\beta^{r}
+ \frac{2\sigma}{3}\hat{\Delta}^{r}\hat{\nabla}_m\beta^{m} \nonumber \\
	& + 2\alpha A_{a}\hat{\Delta}^{r} - 8\pi j_{r} \frac{\xi \alpha}{a} \ ,\\
	L_{2(\Psi)} &= \partial_{r}(\alpha\Pi) \ , \\
	L_{3(\Psi)} &= \beta^{r} \partial_{r}\Psi  + \Psi\partial_{r}\beta^{r} \ , \\
	L_{2(\Pi)} &= \frac{\alpha}{ae^{4\chi}}[\partial_{r}\Psi +\Psi\biggl(\frac{2}{r}-\frac{\partial_{r}a}{2a}+\frac{\partial{r}b}{b}+2\partial_{r}\chi\biggl)\biggl]\nonumber\\
&+\frac{\Psi}{ae^{4\chi}}\partial_{r}\alpha - \alpha \mu^{2}\Phi \,, \\
	L_{3(\Pi)} &= \beta^{r}\partial_{r}\Pi +\alpha K\Pi \ .
\end{align}
Finally, $B^r$ is evolved partially implicitly, using the updated values of 
$\hat{\Delta}^{r}$, i.e., 
$\displaystyle L_{2(B^r)} = \frac{3}{4}\partial_{t}\hat{\Delta}^{r}$ and 
$L_{3(B^r)} = 0$.


\bibliography{num-rel}

\end{document}